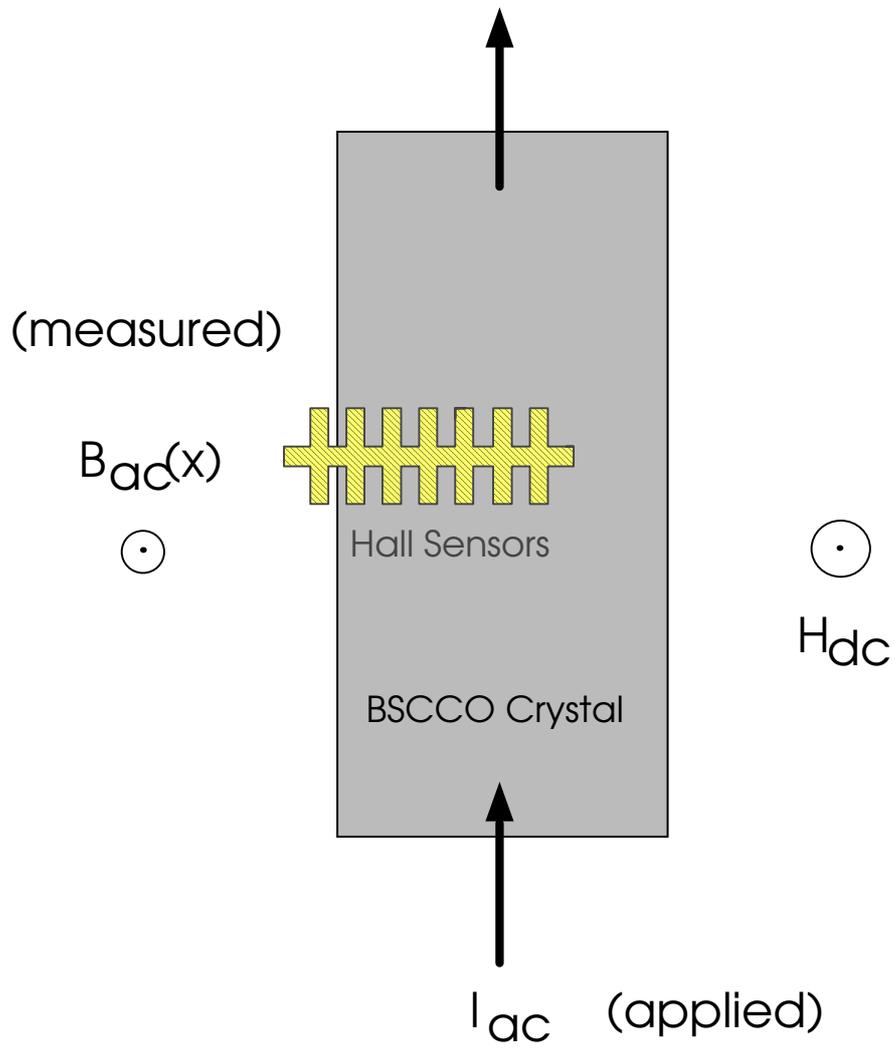

Fuchs et al. Figure 1

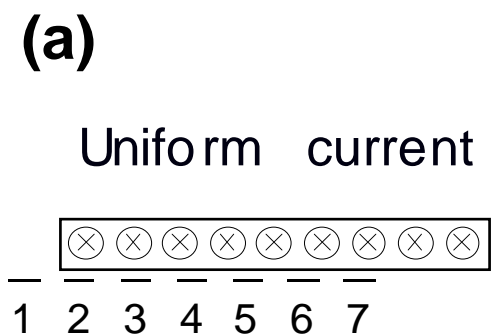
(a) Uniform current

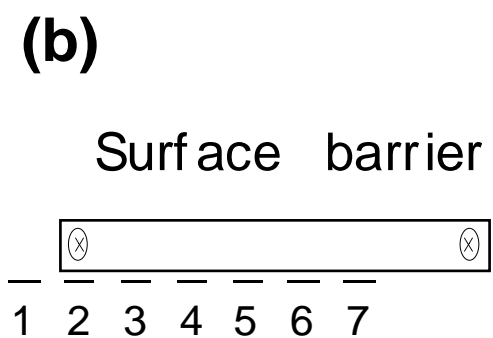
(b) Surface barrier

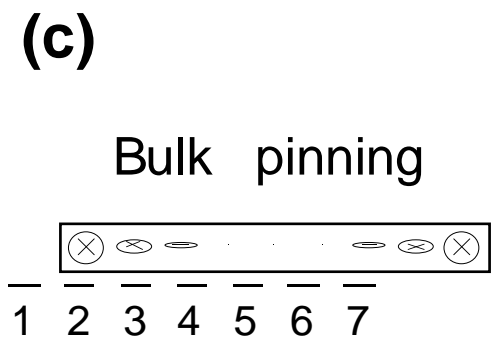
(c) Bulk pinning

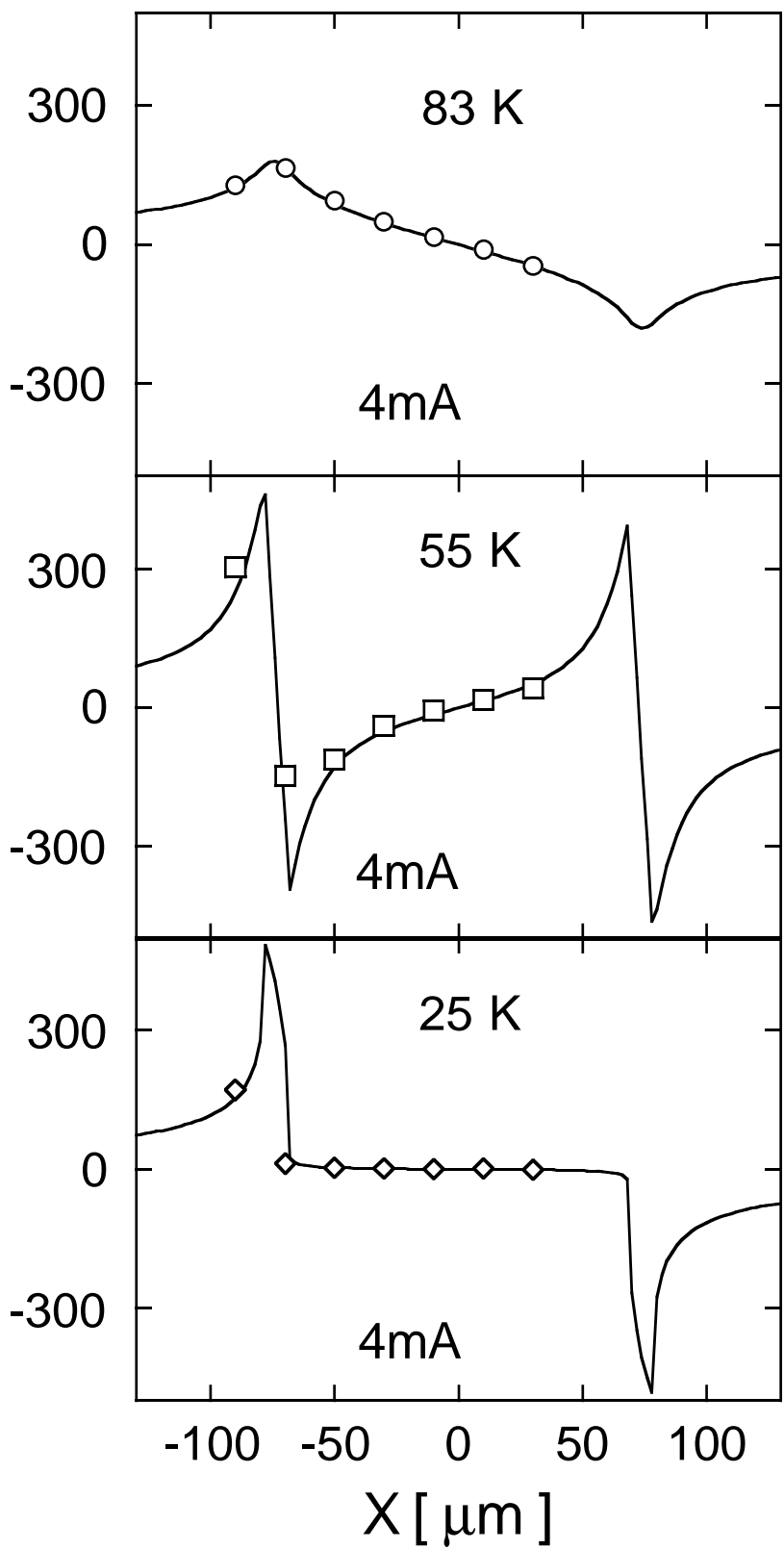

Fuchs et al. Figure 2

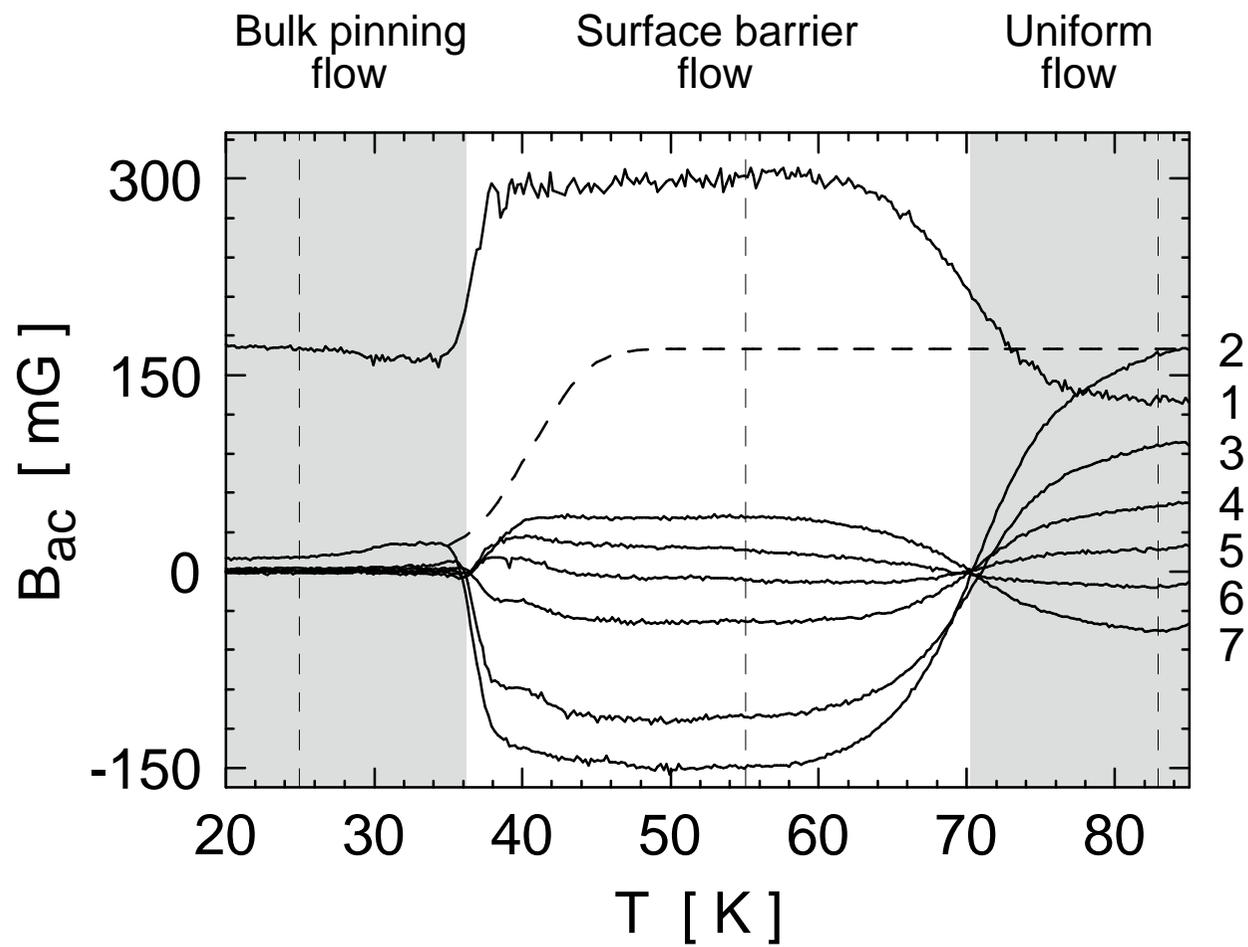

Figure 3
Fuchs et al.

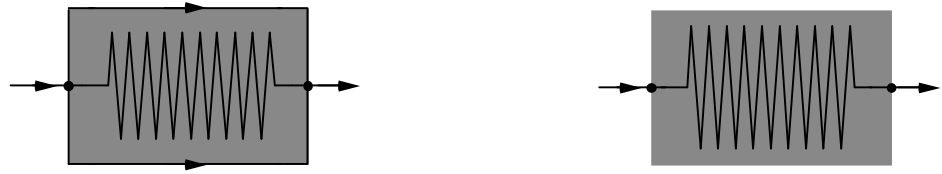
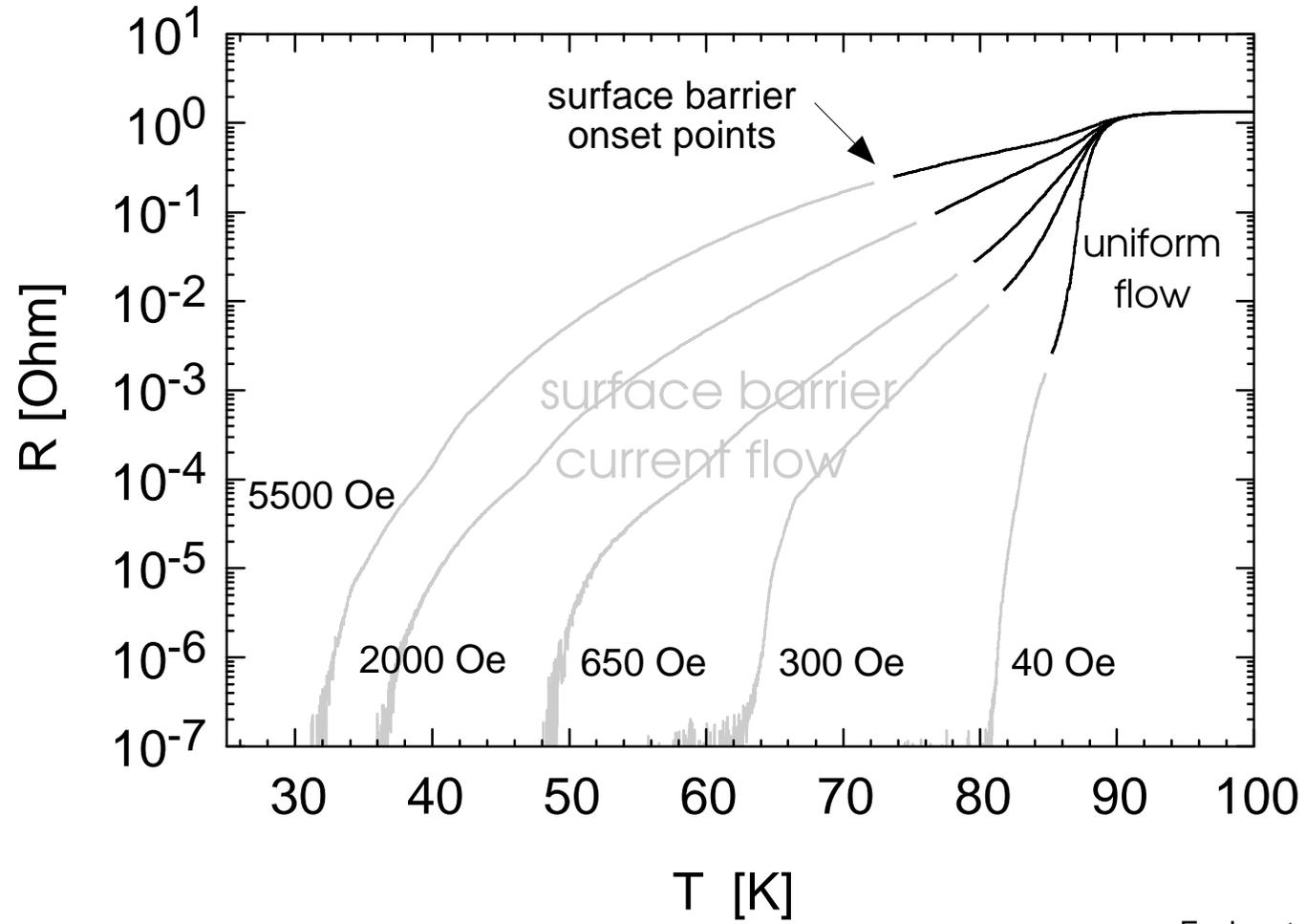

Fuchs et al. Figure 4

# Where does the transport current flow in $Bi_2Sr_2CaCu_2O_8$ crystals?

Dan T. Fuchs[1], Eli Zeldov[1], Michael Rappaport[2], Tsuyoshi Tamegai[3], Shuuichi Ooi[3], Hadas Shtrikman[1]

[1]Department of Condensed Matter Physics, The Weizmann Institute of Science, Rehovot 76100, Israel
[2]Physics Services, The Weizmann Institute of Science, Rehovot 76100, Israel
[3]Department of Applied Physics, The University of Tokyo, Hongo, Bunkyo-ku, Tokyo, 113, Japan

**One of the most common investigation techniques of type-II superconductors is the transport measurement in which electrical current is applied to a sample and the corresponding resistance is measured as a function of temperature and magnetic field. At temperatures well below the critical temperature $T_c$, the resistance of a superconductor is usually immeasurably low. At elevated temperatures and fields, however, in the so-called vortex liquid phase, a substantial linear resistance is observed[1]. In this dissipative state, which in anisotropic high-temperature superconductors like $Bi_2Sr_2CaCu_2O_8$ may occupy most of the mixed state phase diagram, the transport current is usually assumed to flow uniformly across the sample as it does in a normal metal. In order to test this assumption, we have devised a measurement approach which allows determination of the flow pattern of the transport current across the sample. The surprising result is that in $Bi_2Sr_2CaCu_2O_8$ crystals most of the current flows at the edges of the sample rather than in the bulk, even in the highly resistive state, due to the presence of strong surface barriers. This finding has major implications for the interpretation of the existing resistive data and may be of importance for the development of high-temperature superconducting wires and tapes.**

The experiments were carried out on several $Bi_2Sr_2CaCu_2O_8$ (BSCCO) crystals ($T_c \cong$ 88 K) with typical dimensions of 1.5 mm × 0.15 mm × 0.01 mm. The crystals were attached directly to the surface of a linear Hall-sensor array consisting of seven 10 × 10 $\mu m^2$ sensors as shown schematically in Fig. 1. A dc magnetic field, $H_{dc}$, was applied perpendicular to the surface (parallel to the c-axis of the crystal). A low frequency (65 to 400 Hz) ac current, $I_{ac}$, in the range of 0.4 to 10 mA (rms), was applied to the sample using Au contacts. The findings described below are independent of the frequency and the amplitude of $I_{ac}$ and hence only 4 and 10mA at 72.8 Hz data are presented. In this study, we have used a sensitive ac technique to measure the field profile $B_{ac}(x)$ across the sample, where $B_{ac}(x)$ is the self-induced field of the transport current $I_{ac}$. This method differs from the measurements of $B_{dc}(x)$ induced by $H_{dc}$ which provide valuable information about the local magnetization or susceptibility of the sample[2,3]. The present study requires high resolution since $B_{ac}$ is of the order of 0.1 G as compared to the typical $B_{dc}$ of 1000 G. The use of the ac mode provides the necessary separation between the field $B_{ac}$ due to transport current, and the field $B_{dc}$ due to the applied field. The transport current distribution across the sample is then determined from the experimental $B_{ac}(x)$ profiles.

In a highly dissipative state, the current is expected to flow uniformly across the sample like in a normal conductor, as shown on the left in Fig. 2a. In this case, the resulting vertical component of the self-induced field is given by the solid curve on the right in Fig. 2a, as calculated using the Biot-Savart law[4]. At low temperatures, on the other hand, material disorder pins the vortices and prevents their motion, resulting in a finite critical current. In this case the transport current is expected to flow in a way similar to the case of the Meissner state where $B_{ac}(x)$ is expelled from the sample[5,6] as shown in Fig. 2c. As the superconductor is cooled down in the presence of $H_{dc}$, a monotonic increase of the critical current is expected and therefore a gradual transition from $B_{ac}$ of Fig. 2a to Fig. 2c should occur[5,6].

Yet there is another important mechanism, the Bean-Livingston surface barrier[7], which is usually not taken into account in transport studies. [In this paper we use the term surface barrier to refer to both the Bean-Livingston and a similar geometrical barrier mechanism[8]. Since the presented data are mainly for $H_{dc} > H_{c1}(T)$ (the lower critical field), the contribution of the geometrical barrier is negligible in this case]. In order to enter or exit a superconductor, a vortex has to overcome a potential barrier at the surface. This barrier is the result of competition between two forces acting on the vortex near a parallel surface: an inward force due to the presence of the shielding currents, and an outward force due to the attraction between the vortex and its fictitious mirror image outside the sample. In the presence of significant surface barrier, the transport current flows at the edges of the sample, where the vortices enter and leave the superconductor, in order to drive the vortices over the barrier. $B_{ac}(x)$, calculated assuming the entire current flows at the edges, is shown by the solid curve in Fig. 2b. This field profile is quite different from that of the uniform flow in Fig. 2a. In particular, the signs of $B_{ac}(x)$ within the sample are opposite.

It has been shown that surface barriers have an important contribution to the hysteretic magnetization of clean high-$T_c$ superconductors[9,10,11]. These barriers govern the magnetization below the so-called irreversibility line. Due to practical sensitivity limitations, transport measurements are usually carried out above this irreversibility line where the magnetization is reversible and hence surface effects are expected to be of no importance. The effect of surface barriers on transport measurements has been considered theoretically[12], and some studies have suggested that surface barriers may be of significance[13], in particular as a source of voltage noise[14,15]. The flow patterns of the transport current have been tested in several investigations on thin films and tapes[16,17,18,19], however, these studies have focused on the critical state behaviour and sample inhomogeneities rather than on surface barriers.

Figure 3 shows the self-field $B_{ac}$ as measured by the sensors upon cooling the BSCCO crystal in $H_{dc} = 0.1$ T and $I_{ac} = 4$ mA. The curves are labelled by the sensor number as indicated in Fig. 2. At elevated temperatures, $B_{ac}(x)$ decreases monotonically across the sample (sensors 2 through 7) due to a uniform transport current. The measured field profile at 83 K is shown in Fig. 2a by the open symbols along with the calculated field profile for 4 mA uniform current shown by the solid curve. The clear fit indicates that at high temperatures the transport current flows uniformly across the width of the crystal. At low temperatures, T<36 K in Fig. 3, $B_{ac}(x)$ is finite only outside the bulk of the sample (sensor 1 and sensor 2 which is close to the edge) and is zero inside the sample. This behaviour indicates strong bulk pinning and a finite critical current. Figure 2c shows the measured field profile at 25 K and the corresponding calculated curve for this case. At intermediate temperatures, if the surface barrier is neglected, one expects a gradual crossover from a uniform current flow to bulk pinning as indicated by the dashed line for sensor 2 in Fig. 3. However, the measured $B_{ac}$ is drastically different and displays a negative signal which cannot be explained by any model based on bulk vortex properties. This inversion of the polarity of $B_{ac}$ at T = 70 K for all sensors within the sample is the result of current flow at the edges of the sample due to the presence of surface barriers. The measured field profile at T = 55 K is shown in Fig. 2b. The solid curve is calculated for the 4 mA current flowing entirely at the edges of the sample. The clear agreement with the data indicates that the surface barrier completely dominates the vortex dynamics, and as a result, practically the entire current, within experimental resolution of few percent, flows at the edges of the sample. Figure 3 shows that at high temperatures the current flows uniformly, then crosses over to a complete surface flow over a wide range of temperatures, and finally changes to bulk-pinning-like flow only at low temperatures.

The above results have major implications for the interpretation of transport measurements. Figure 4 shows the resistance transition in the same BSCCO crystal at various applied fields. Such measurements are usually interpreted in terms of bulk resistivity caused by the viscous vortex flow, pinning, etc. Apparently, only the very top region of our data, where the current flows uniformly in the bulk (dark lines), reflects bulk vortex properties. Most of the data shown by the gray lines, however, do not reflect bulk vortex properties but rather the properties of the surface barrier. The situation can be roughly represented by the two equivalent circuits displayed at the top of the figure. At elevated temperatures, the sample can be regarded as a uniform resistor. However at lower temperatures (gray lines) the surface barriers act as low resistance shunts, and hence practically all the current flows at the edges. As a result, the total measured resistance is that of the surface barrier rather than that of the bulk.

An alternative way of describing the process is by noting that the measured resistance always reflects the flow rate of the vortices across the sample. However, the flow rate, rather than being determined by bulk vortex pinning and viscous drag, is determined by the hopping rate over the surface barrier[12]. This activation process is the bottleneck impeding vortex motion. Vortex flow rate over the barrier has to be equal to that in the bulk. As a result practically all of the current flows at the edges in order to provide the large force required to overcome the barrier, while only a very small fraction of the current flows in the bulk in order to overcome the small bulk viscous drag. The apparent measured resistance is thus orders of magnitude lower than the true bulk resistance. In other words, the vortices in the bulk are much more mobile, and the corresponding viscous drag coefficient and the pinning force are much smaller, than what is usually deduced from transport measurements[20,21,22,23].

In addition, resistance of BSCCO crystals commonly displays thermally activated Arrhenius behaviour. In view of this work, this is caused by the thermal activation over the surface barrier[12] and does not reflect the bulk pinning, as is usually interpreted[20,21]. Another interesting observation is the recently reported sharp drop in the resistance at the first-order vortex lattice melting transition in BSCCO[24,25,23], which is seen for example, in the 300 Oe data in Fig. 4 at about 65 K. Such a drop is usually ascribed to the sharp onset of bulk pinning upon solidification of the lattice[26,27]. Analysis of the temperature dependence of the current distribution at 300 Oe indicates that even this feature mainly reflects a sharp change in the surface rather than bulk properties since most of the current still flows at the edges of the crystal and the vortices are unpinned both above and below the melting transition. This does not mean that the bulk properties do not change, but rather that the transport measurements do not probe the true bulk vortex dynamics at the transition. Finally several magneto-optical investigations have shown that in BSCCO tapes and wires the current distribution often peaks at the interface between the superconducting filaments with the Ag sheath rather than within the bulk of the filaments[19]. It is thus possible that the surface barriers are of significant importance also to the current carrying capability of the commercial wires in a way similar to the behaviour of the single crystals.

Figure Captions

Figure 1

Schematic top view of $Bi_2Sr_2CaCu_2O_8$ crystal (1.5 mm × 0.15 mm × 0.01 mm) attached to an array of seven GaAs/AlGaAs two-dimensional electron-gas Hall sensors. The sensors have an active area of $10 \times 10$ μm$^2$ and are 10μm apart. There is one sensor outside the sample and the other six span more than half of the sample width. Magnetic field $H_{dc}$ is applied perpendicular to the sample surface. An ac current $I_{ac}$ is applied to the crystal through electrical contacts. The sensors are used to probe the perpendicular component of the self-induced ac field, $B_{ac}$, generated by the ac current.

Figure 2

Schematic cross section of the sample with the attached Hall sensors (left) and the corresponding field profiles $B_{ac}(x)$ at three temperatures (right). Open symbols are the self-field measured by the sensors ($H_{dc}$ = 1000G, $I_{ac}$ = 4 mA), and the solid lines are the calculated field profiles for 4 mA current using the Biot-Savart law. (a) At elevated temperatures, the current flows uniformly across the sample. (b) At intermediate temperatures practically all the current flows at the edges of the superconductor due to surface barriers. In this case the field inside the sample has an inverted profile relative to the uniform-current-flow case. (c) At low temperatures, strong bulk pinning is present which results in zero $B_{ac}$ within the sample.

Figure 3

Self-induced field $B_{ac}(x)$ generated by 4 mA ac current as a function of the temperature during cooling of the $Bi_2Sr_2CaCu_2O_8$ crystal in $H_{dc}$ field of 0.1 Tesla. The curves are labelled according to the sensor numbers as indicated in Figure 2. The vertical dashed lines mark the temperatures for which the $B_{ac}$ values are plotted as a function of the sensor location in Figure 2. At elevated temperatures the current flows rather uniformly across the sample (83 K profile in Fig. 2a). Surface barriers dominate the vortex dynamics over a wide range of intermediate temperatures where the currents flows at the edges of the sample (55 K profile in Fig. 2b). At the crossing point at 70 K, half of the current flows in the bulk and half at the edges causing cancellation of the field in the central part of the sample (superposition of profiles (a) and (b) of Fig. 2). At low temperatures, strong bulk pinning prevents vortex motion, resulting in zero $B_{ac}$ in the sample. The dashed line indicates the expected field at sensor 2 if the current flow would be governed by the bulk vortex properties as commonly assumed. The observed inversion of the field profile is caused by the dominant role of surface barriers.

Figure 4

Resistance of $Bi_2Sr_2CaCu_2O_8$ crystal as a function of temperature at the indicated applied fields ($I_{ac}$=10mA). Only in the region of the solid curves the current flows uniformly across the sample (at least 90% of the current). In the gray region most of the current flows at the edges of the sample due to strong surface barriers. The top left diagram illustrates how the surface barriers shunt the current flow in a large part of phase diagram, (surface barrier flow region in Fig. 3), and as a result the measured resistance may be orders of magnitude lower than the true bulk resistance of the superconductor. The top right diagram illustrates the uniform flow case where the surface barriers are unimportant.

**Acknowledgments**: We thank M. Konczykowski, R. Doyle, V. Kogan, M. McElfresh, and A. Koshelev for valuable discussions. This work was supported by the Israel Science Foundation, by German-Israeli Foundation for Scientific Research and Development (GIF), by MINERVA foundation, Munich, Germany, by the Alhadeff Research Award, and by the Grant-in-Aid for Scientific Research from the Ministry of Education, Science, Sports and Culture, Japan.

Correspondence and requests for materials should be addressed to D. T. F. (email : fndandan@wis.weizmann.ac.il)